\documentclass[acus]{JAC2003}

\usepackage{epstopdf}
\usepackage{graphicx}
\usepackage{booktabs}

\begin{document}
\title{The potential for a high power FFAG proton driver for ADS }

\author{S. L. Sheehy\thanks{suzie.sheehy@stfc.ac.uk}, STFC/RAL/ASTeC, Didcot, Oxon, UK,\\
C. Johnstone, PAC, Batavia, Illinois, USA,\\
R. Barlow, University of Huddersfield, UK, \\
A. Adelmann, PSI, Villigen, Switzerland.}

\maketitle

\begin{abstract}
Fixed-field alternating gradient accelerators are promising candidates for next-generation 10 MW-class high power proton drivers. Recent advances in lattice design of non-scaling FFAGs have progressed toward both isochronicity and chromatic correction. The resulting 1~GeV non-scaling FFAG design may be able to support a continuous (CW) beam with far lower peak current than the pulsed alternative. A 6-cell non-scaling FFAG design is described and recent work in modeling 3D space charge using the OPAL framework is presented, including fixed energy studies and beam dynamics with fast acceleration in the so-called Ôserpentine channelÕ.  
   \end{abstract}

\section{INTRODUCTION}
\subsection{FFAG design}

In the 2010 US White Paper on accelerator technologies for Accelerator Driven Systems~\cite{whitepaper}, FFAG accelerators were described as ``promising [...] albeit without the capability for true CW operation''. Since this time the design of non-scaling FFAGs has evolved to cover the spectrum between the scaling FFAG and the linear non-scaling FFAG~\cite{emma, sheehy10, johnstone11}. Two advancements in particular make their use as high power proton drivers a potentially attractive option: minimising the tune variation so they avoid crossing betatron resonances and moving towards isochronous orbits. Incorporating both of these ideas in a single design would produce a compact fixed-field, fixed-RF CW accelerator design with stable tunes.

\begin{figure}[h!]
\centering
\includegraphics[width=0.7\linewidth]{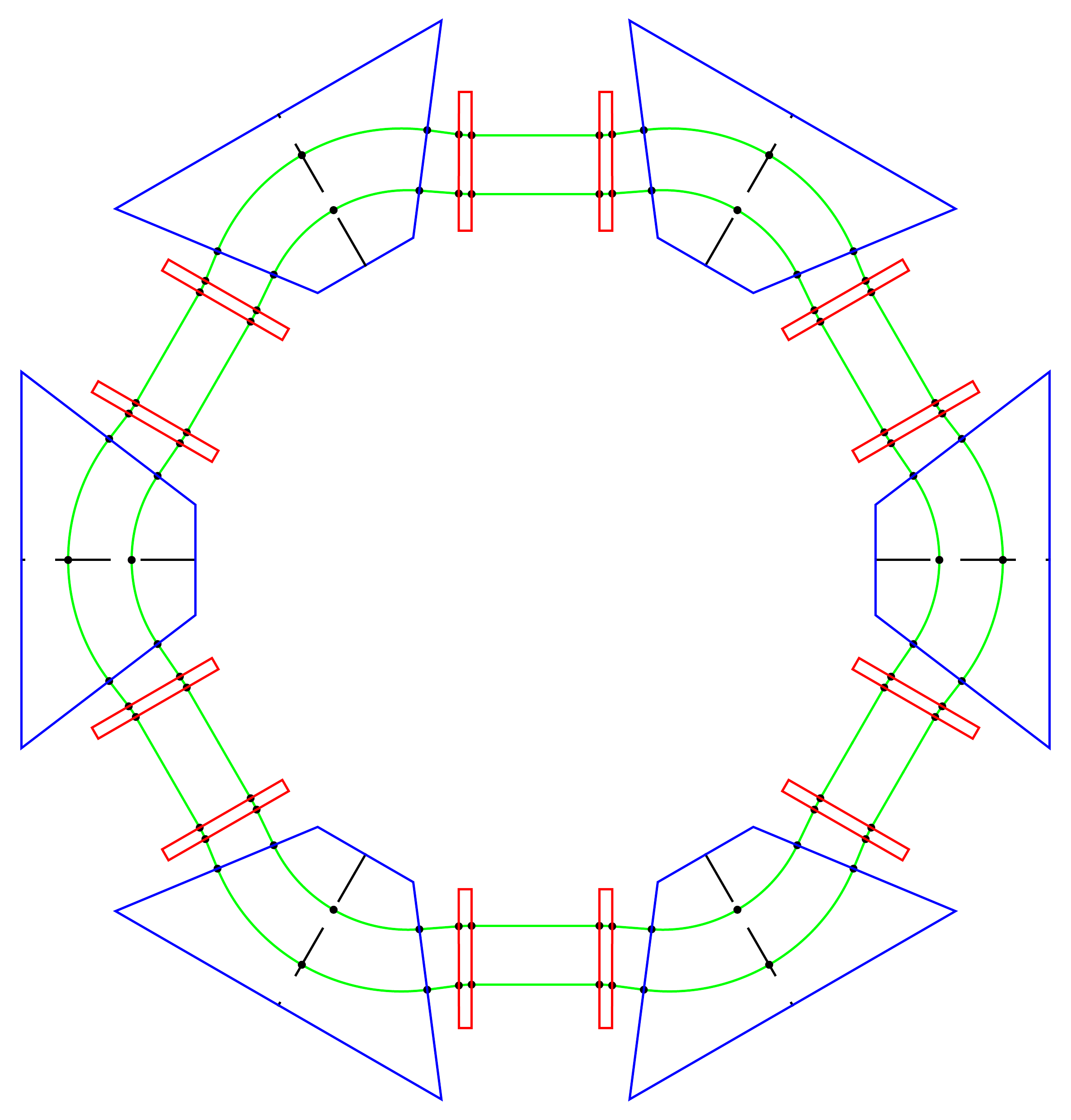}
\caption{Layout of 6 cell DFD lattice design with green traces showing a low and intermediate energy orbits.\label{figure1}}
\end{figure}

The lattice design discussed here is aimed towards achieving these two things simultaneously; that is, achieving nearly isochronous orbits while minimising the tune variation (or chromaticity). A recent six cell FFAG lattice design is used as a representative model in order to study the relevant beam dynamics. It consists of six triplet cells in a DFD configuration with long drift sections of approximately two metres as shown in Fig.~\ref{figure1}. The design procedure is outlined in more detail in Ref.~\cite{johnstone_aip10, johnstone_cycl10} and parameters are given in Table~\ref{table1}.

The lattice incorporates wedge-shaped magnets with a radial magnetic field dependence in order to produce stable tunes and achieves a time-of-flight variation that is within $\pm 0.5\%$ over the full acceleration range. The basic transverse dynamics have been verified three-fold; in COSY-infinity~\cite{cosy} as part of the original design, using the tracking code ZGOUBI~\cite{zgoubi, sheehy11_2} and here using particle tracking in the OPAL framework~\cite{opal}.

\begin{table*}
\begin{small}
\begin{center}
\caption{General parameters, 6-cell 1 GeV FFAG.}
\begin{tabular}{cccc}
  \toprule
  Parameter	& 330 MeV	& 500 MeV	& 1000 MeV \\
Avg. Radius [m] & 5.498 &	6.087 &	7.086 \\
$\nu_x$/$\nu_y$  (cell) &	0.297/0.196	& 0.313/0.206	& 0.367/0.235 \\
Field F/D [T] &	1.7/$-$0.1 & 1.8/$-$1.9 & 1.9/$-$3.8 \\
Magnet Size F/D [m]& 1.96/0.20 &	2.79/0.20 &	4.09/0.20 \\
\end{tabular}
\label{table1}
\end{center}
\end{small}
\end{table*}

\begin{figure}
\centering
\includegraphics[width=0.8\linewidth]{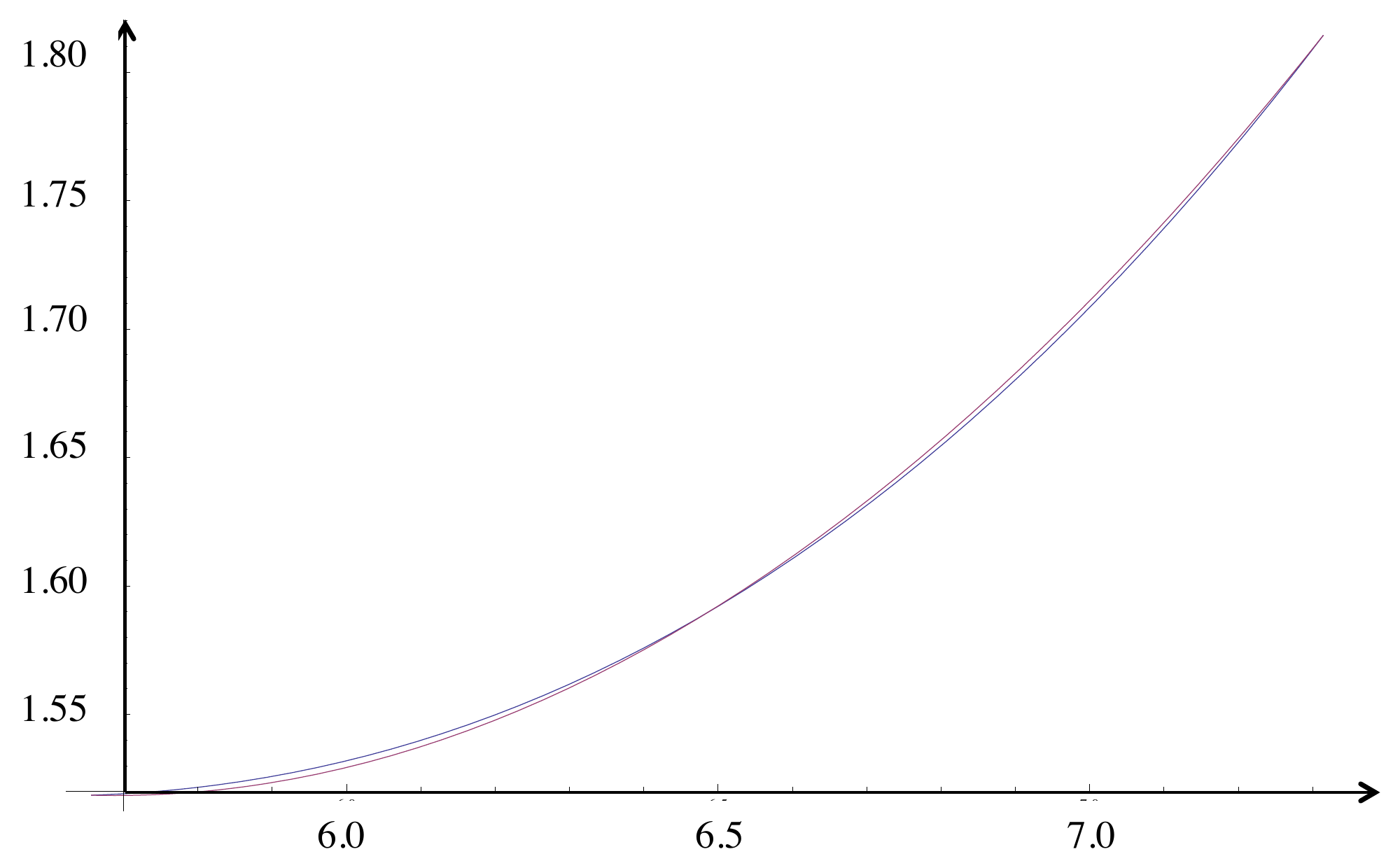}
\caption{Magnetic field profile [T] with radius [m] in the focusing (F) magnets\label{figure2a}}
\end{figure}

\begin{figure}
\centering
\includegraphics[width=0.8\linewidth]{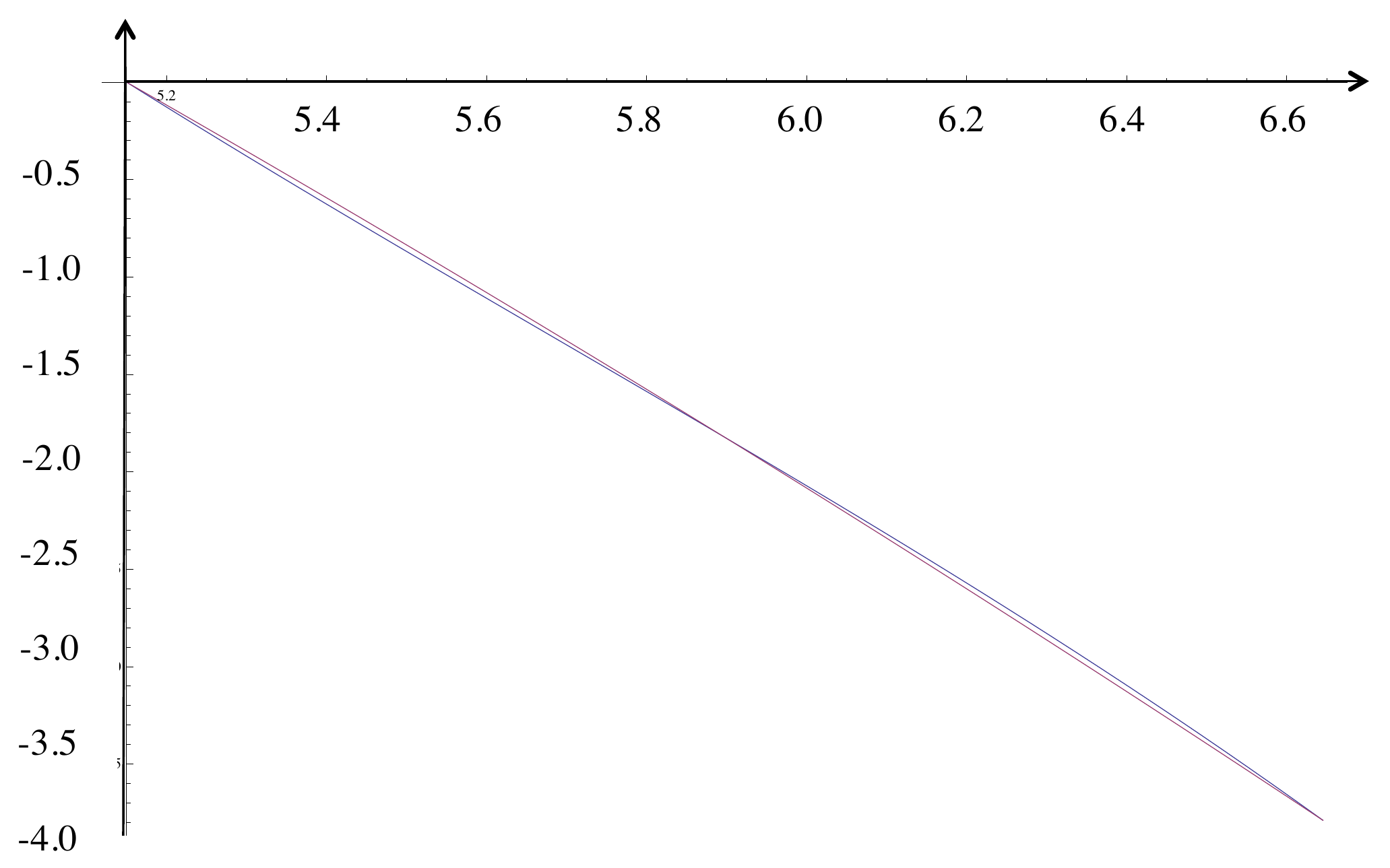}
\caption{Magnetic field profile [T] with radius [m] in the defocusing (D) magnets\label{figure2b}}
\end{figure}

The magnetic field has a substantial variation with radius, as shown in Figs.~\ref{figure2a} and~\ref{figure2b}. A detailed mid-plane magnetic field map is generated as part of the design procedure, which is then imported into the OPAL framework for beam dynamics studies.

\subsection{Basic dynamics}

\begin{figure}
\centering
\includegraphics[width=1.0\linewidth]{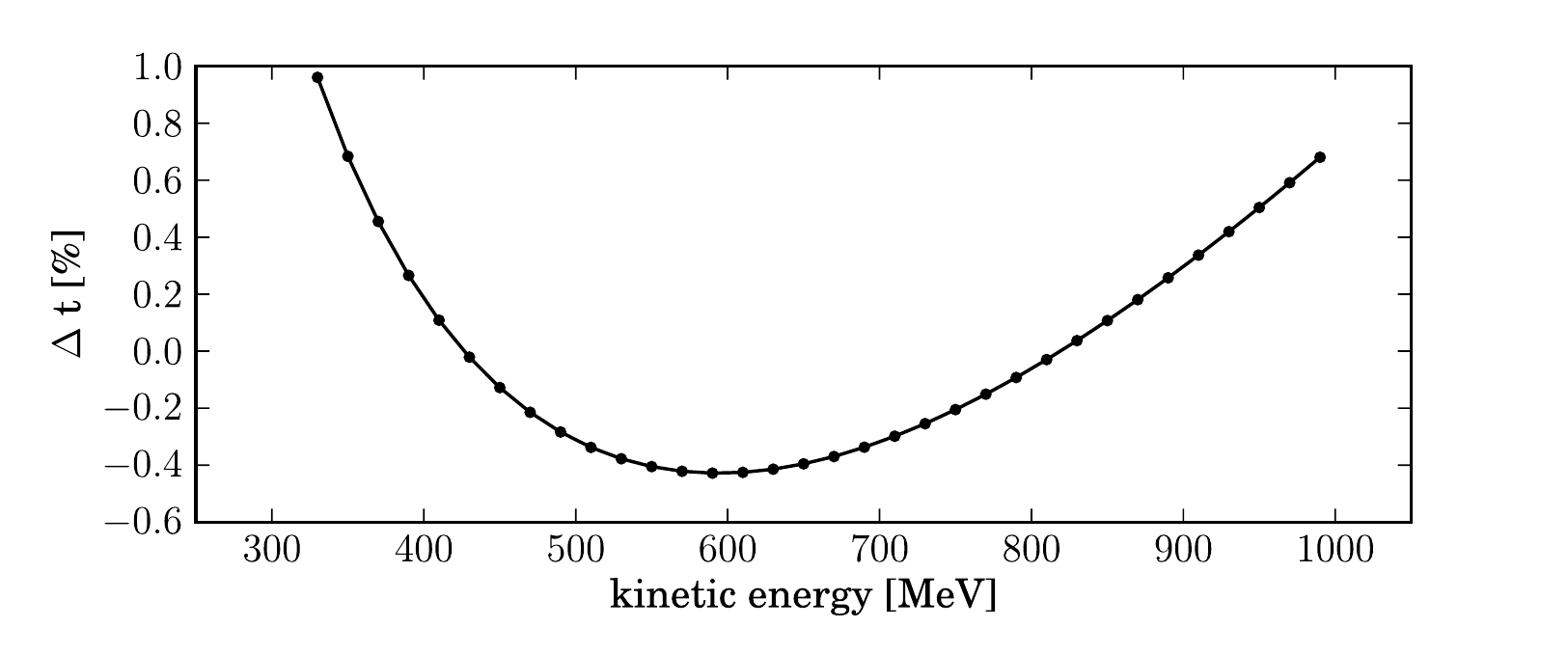}
\caption{Variation of time of flight with kinetic energy from OPAL in 20 MeV steps. \label{figure3}}
\end{figure}

The design is not perfectly isochronous with a time-of-flight variation with energy of up to $\pm1\%$. The closed orbit time-of-flight variation with energy calculated using single particle tracking in OPAL is shown in Fig.~\ref{figure3}.

\begin{figure}
\centering
\includegraphics[width=1.0\linewidth]{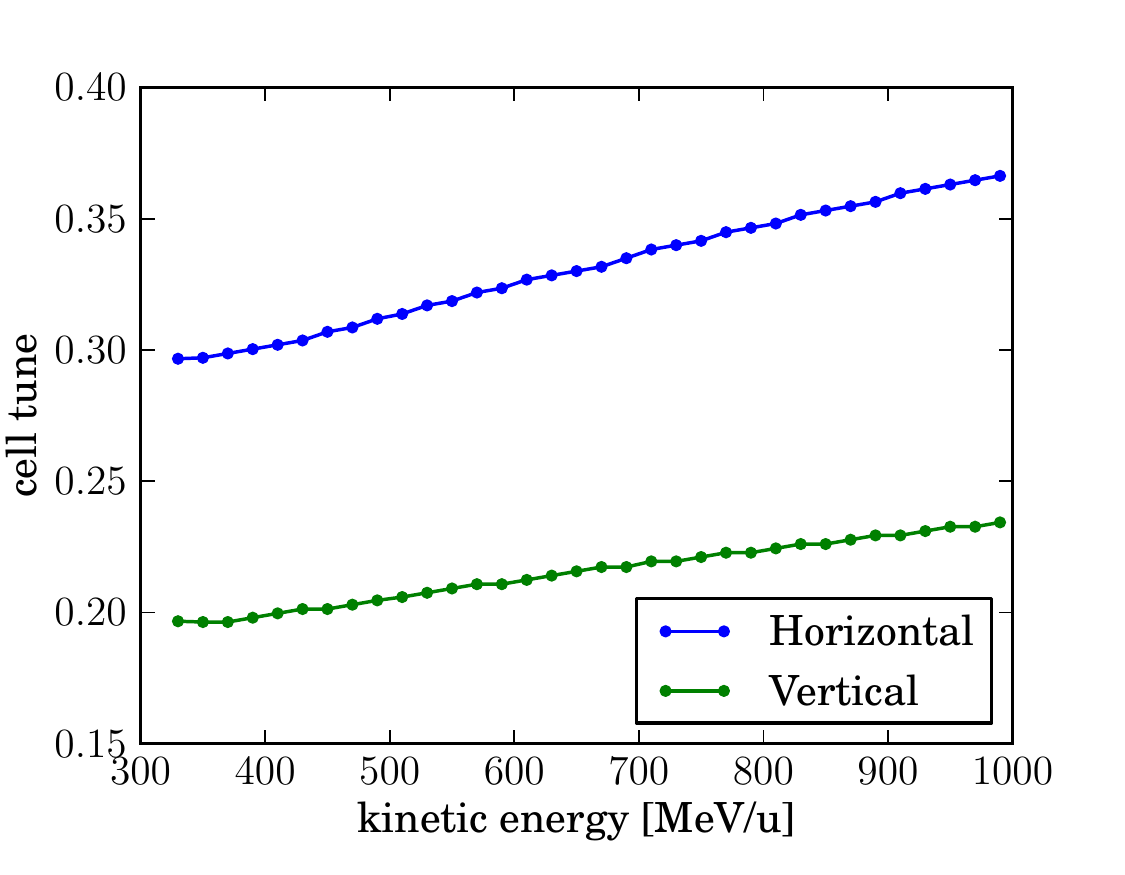}
\caption{Variation of betatron tunes with kinetic energy from OPAL in 20 MeV steps.\label{figure4}}
\end{figure}

The betatron cell tunes are shown in Fig.~\ref{figure4}. Note that in the present design there is a horizontal integer structural resonance at $\nu_{x}\,=\,0.3333$ or $Q_{x}\,=\,2$ at around 690\,MeV. In future the design should be optimised to avoid this.
 
\begin{figure}
\centering
\includegraphics[width=1.0\linewidth]{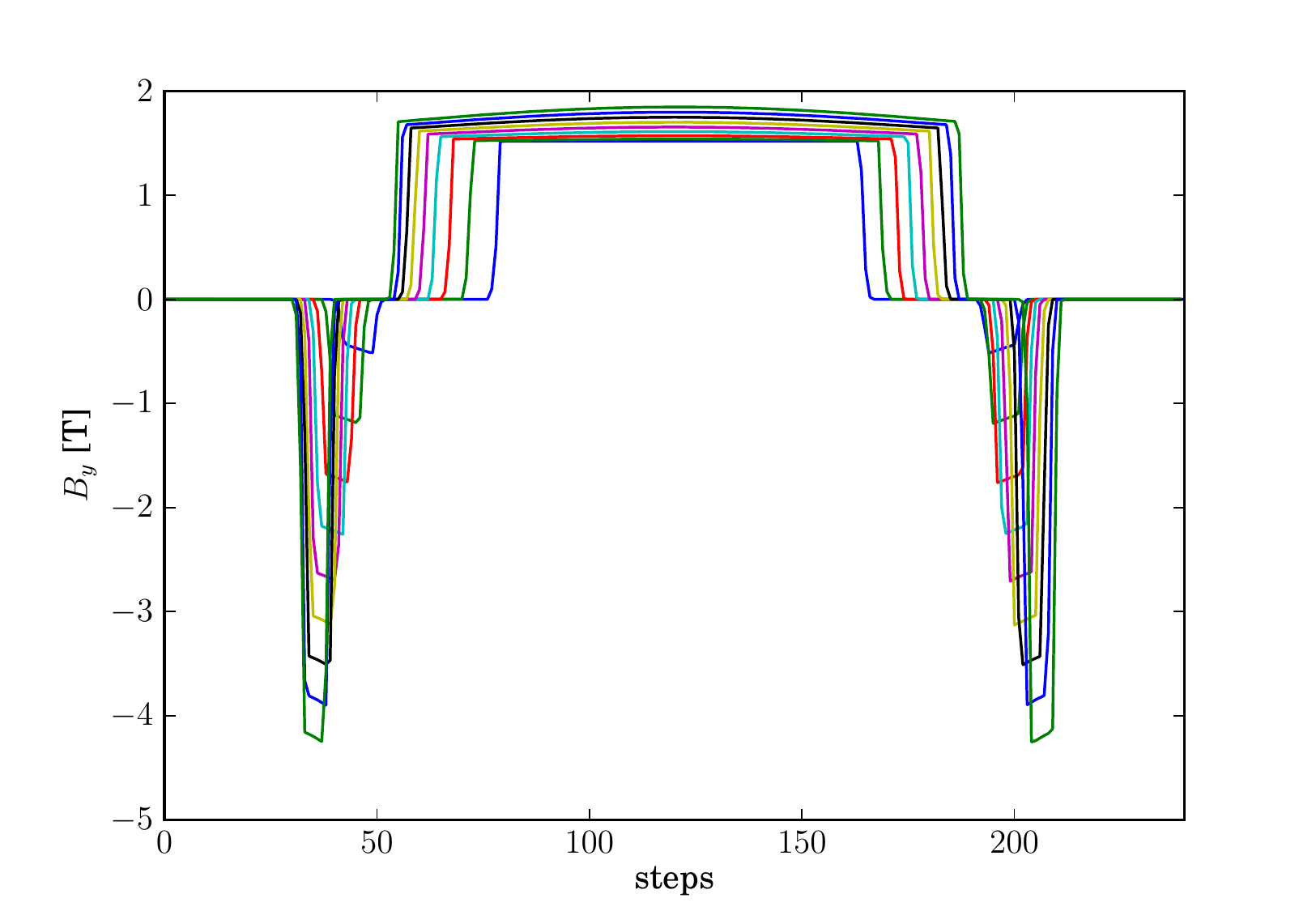}
\caption{Vertical magnetic field experienced by the particles along closed orbits in 80 MeV steps from 330 MeV to 970 MeV.\label{figure5}}
\end{figure}

One item of particular interest is the magnetic field experienced by particles of different energies as they travel along their respective closed orbits, as this defines the actual required magnetic field. This is shown in Fig.~\ref{figure5} and highlights not only the effect of the field profile with radius, but also that of the wedge shaped magnets. 

\section{ACCELERATION MECHANISM}

In cyclotrons the time of flight variation with energy is constant to a level of around $10^{-4}$ which ensures that the particles remain on crest for sufficient turns to be accelerated efficiently before extraction. A lot of effort is put into ensuring this isochronicity through the use of shimming and adjustment coils. In this FFAG design the isochronicity is not perfect as the time of flight still varies by up to $\pm1\%$ throughout acceleration. This is not sufficient to facilitate cyclotron-like acceleration as the bunch phase will move off-crest and be decelerated before the full energy range is achieved. However, there is another option. 

Upon inspection of the time of flight curve with energy in Fig.~\ref{figure3} the roughly parabolic shape is clear, and this opens up the possibility of using the so-called `serpentine channel' for acceleration, recently demonstrated in the EMMA~\cite{emma} experiment. 

\subsection{Model equations}
\begin{figure}
\centering
\includegraphics[width=0.9\linewidth]{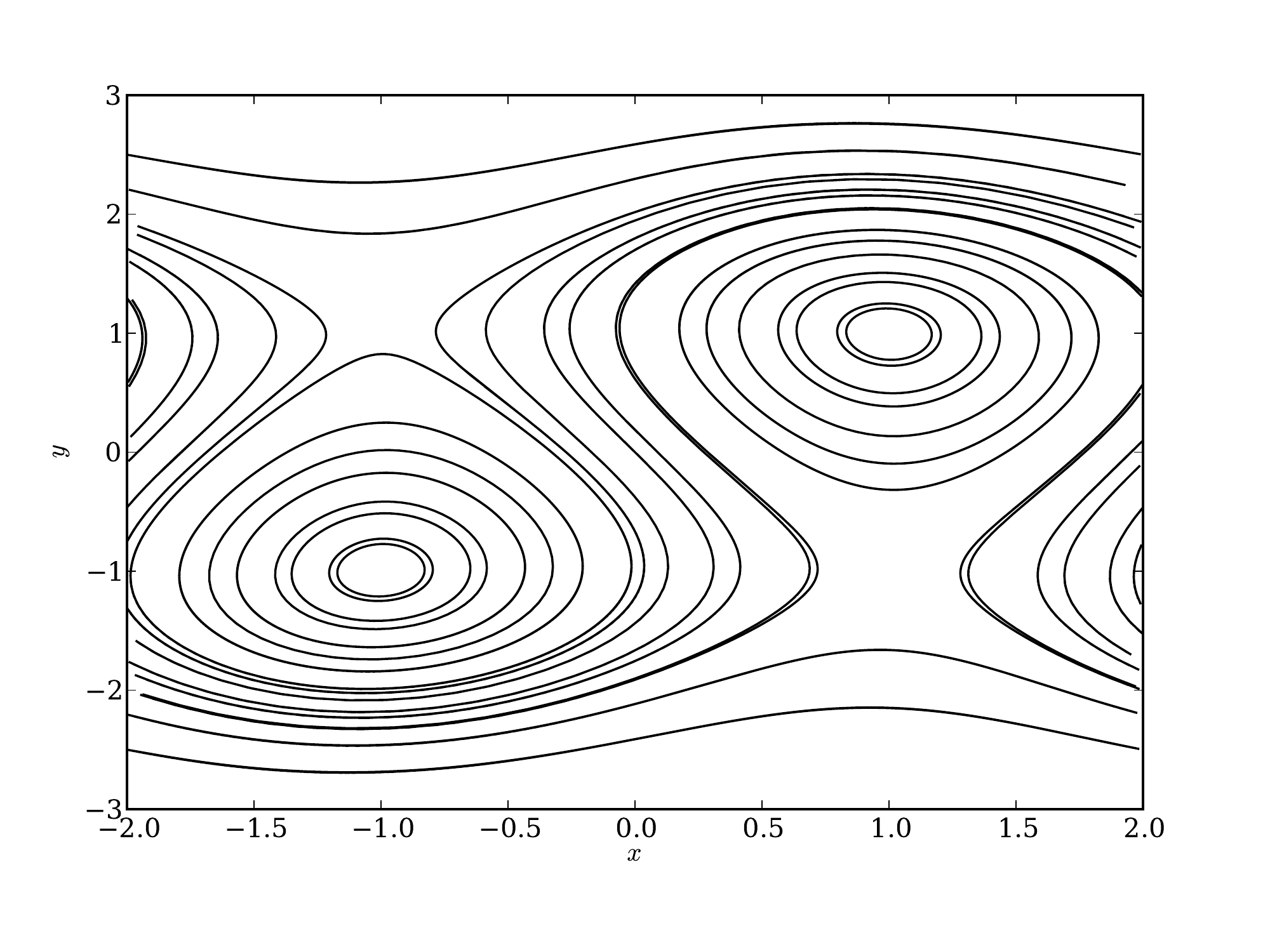}
\caption{Serpentine channel for $\lambda=4/3$ contours plotted for regularly spaced values of x=y.\label{figure6}}
\end{figure}

A system with a parabolic time of flight curve can be approximated by a set of model equations as per Koscielniak \textit{et al.}~\cite{koscielniak04},

\begin{equation}
\frac{dy}{ds}=\lambda\left(\frac{\pi}{2}\right)cos\left(\frac{x\pi}{2}\right)
\end{equation}

\begin{equation}
dx/ds=y^{2}-1,
\end{equation}

where  $y \propto (E-\bar{E})$ and $\bar{E}$ is the mean energy, and x relates to phase. The motion is synchronous at $y=\pm1$ and the direction of the phase slip reverses above/below these values. For values of $\lambda>2/3$ we see (as expected) the emergence of a serpentine libration which creates a `gutter' feature allowing the range of acceleration to exceed that of normal `phase slip' or cyclotron-like acceleration, as in Fig.~\ref{figure6}. Note that in the figure a beam can be injected at $[-1.0, -2.0]$ and extracted at $[1.0, 2.0]$. This leads to two reversals in the direction of phase slip during acceleration and consequently the beam can cross the crest three times before it slips to decelerating values.

To ascertain RF acceleration parameters the cavity type, frequency and harmonic need to be considered. The most basic assumption is a rectangular cavity with no radial variation of voltage, operating at a frequency optimised for serpentine acceleration, in this case 5.795\,MHz. To check the phase acceptance in the FFAG design presented herein (in which the time of flight curve is not perfectly quadratic) single particle tracking in OPAL was undertaken using the serpentine channel from 330\,MeV to 1\,GeV. 

\begin{figure}
\centering
\includegraphics[width=0.8\linewidth]{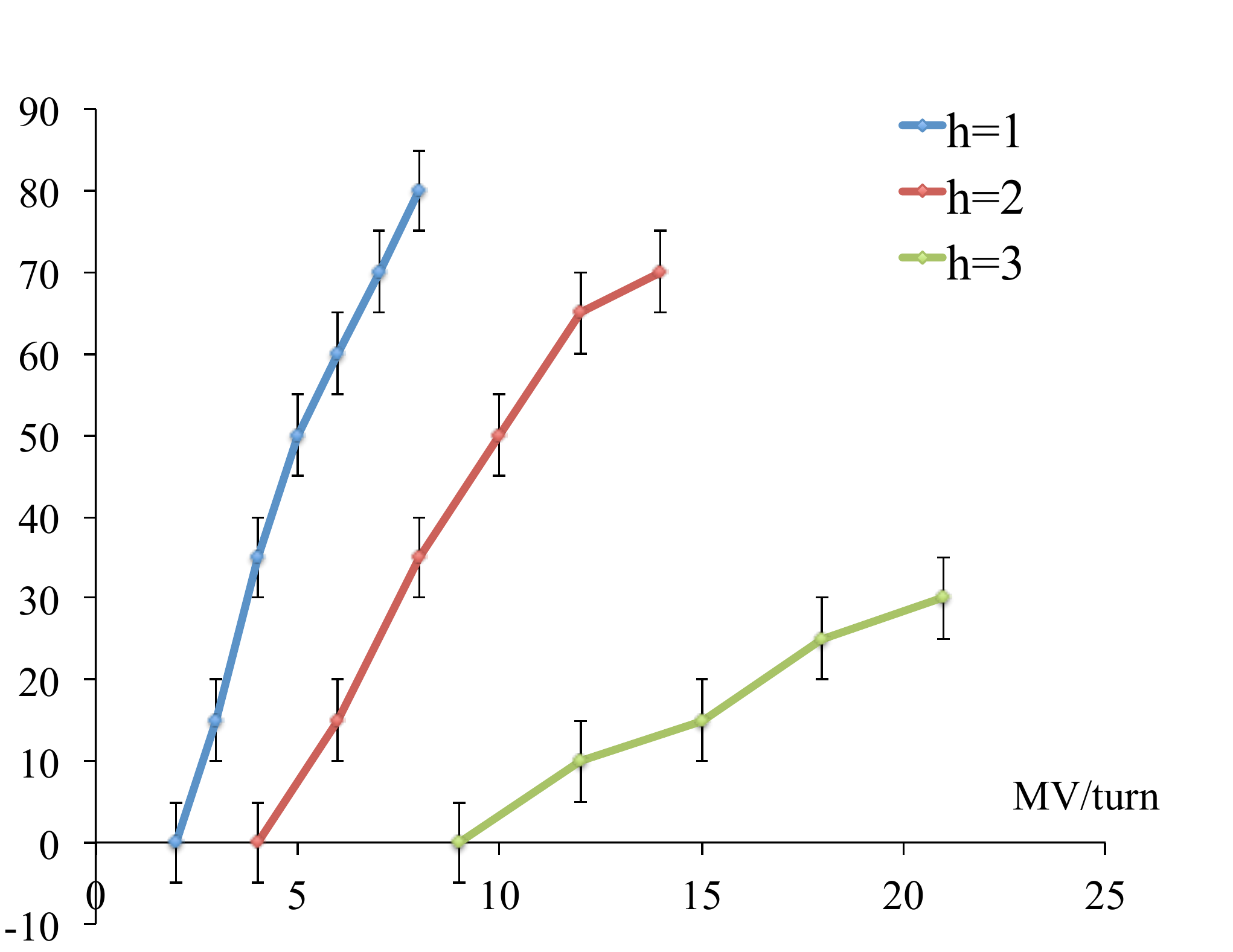}
\caption{Maximum gutter acceptance in degrees for varying voltage per turn and harmonic number.\label{figure7}}
\end{figure}

As expected the particle is able to start at a wide range of RF phase if operating with the first harmonic, but there is a dependence harmonic number. The tracking results are shown in Fig.~\ref{figure7} and the restriction of phase acceptance with large harmonic number will need to be considered in future. It is possible to add higher harmonics to increase the acceptance, however this will require a higher accelerating voltage~\cite{koscielniak04}.

\section{SPACE CHARGE SIMULATIONS}

Having established the basic lattice design, basic stable dynamics and a possible acceleration mechanism, the natural next step is to study the effects of intense beams in such a design. The first step is to match a distribution to the accelerator.

\subsection{Matching}

To achieve zero-current matching at first order, single particle tracking is carried out at a moderate amplitude (up to 10\,mm) around the closed orbit. An ellipse is then fitted to the resulting phase space ellipse to determine the basic twiss parameters, which are then used to define the input distribution in OPAL.

\begin{figure}
\centering
\includegraphics[width=0.8\linewidth]{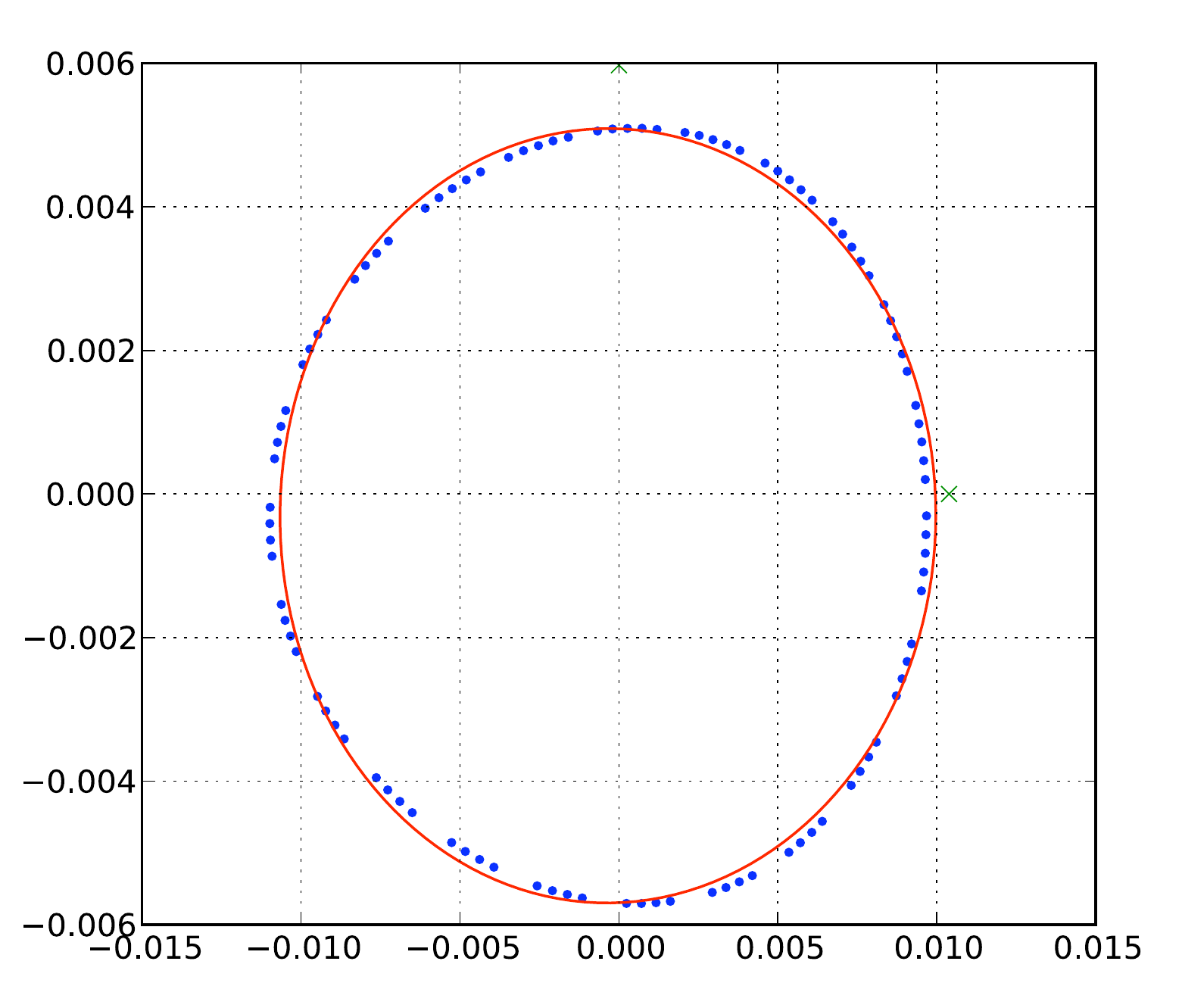}
\caption{Horizontal phase space (position [m] vs conjugate momentum [rad]) with elliptical fit to determine twiss parameters from single particle tracking at 610\,MeV with 10\,mm horizontal amplitude.\label{figure8} }
\end{figure}

In a fairly linear regime this method is quite good at matching with zero-current and the ellipse is well-defined. However when the field is highly non-linear as is the case at around 610\,MeV, the ellipse parametrization is not ideal, as shown in Fig.~\ref{figure8}. This method is also unable to take into account the change in focusing due to space charge effects and the subsequent change in matching parameters.

To the authors' knowledge there is no theoretical method to date to match distributions with space charge in this type of FFAG. A method for matching with space-charge has only recently been developed for isochronous cyclotrons~\cite{baumgarten11}. This method focuses particularly on the transverse-longitudinal coupling present in isochronous cyclotrons, while taking into account small deviations from isochronicity. The applicability of this method to nearly-isochronous FFAGs like the one presented here is currently under investigation.

\subsection{Fixed energy simulations}

The time of flight curve in Fig.~\ref{figure3} has three quite distinct regions; one where the time of flight is rapidly decreasing with energy (330\,MeV - 500\,MeV), a turn-around point where the time-of-flight is roughly constant with energy (approximately isochronous, 500\,MeV to 700\,MeV) and a third where the time of flight is rapidly increasing with energy (700\,MeV to 1\,GeV). While this roughly parabolic time of flight is the very feature that makes this design suitable for serpentine acceleration, it is also interesting to consider the dynamics with space charge in these rather different regimes.

In the `isochronous' regime around the centre of the energy range, it might be expected that this machine acts like an isochronous cyclotron. At a single energy with no RF, there is no longitudinal focusing in an isochronous accelerator so with space charge the beam may be expected to grow in length. To test this a $10\pi$mm.mrad bunch (unnormalised) is tracked using OPAL with no RF voltage applied. 

\begin{figure}
\centering
\includegraphics[width=0.9\linewidth]{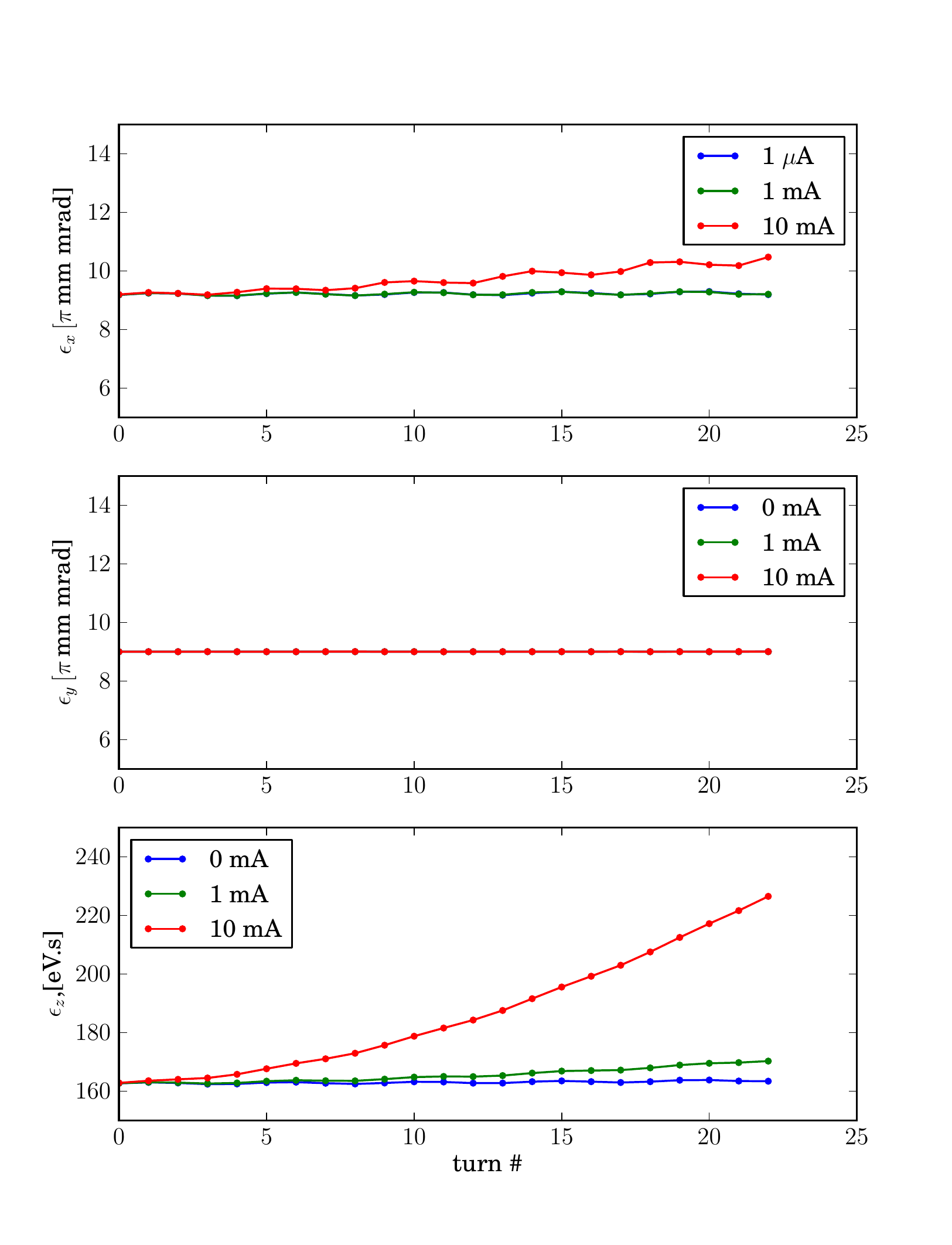}
\caption{Emittance variation at fixed energy at 330\,MeV for a 100\,mm long beam at varying average beam current.}
\end{figure}

\begin{figure}
\centering
\includegraphics[width=0.9\linewidth]{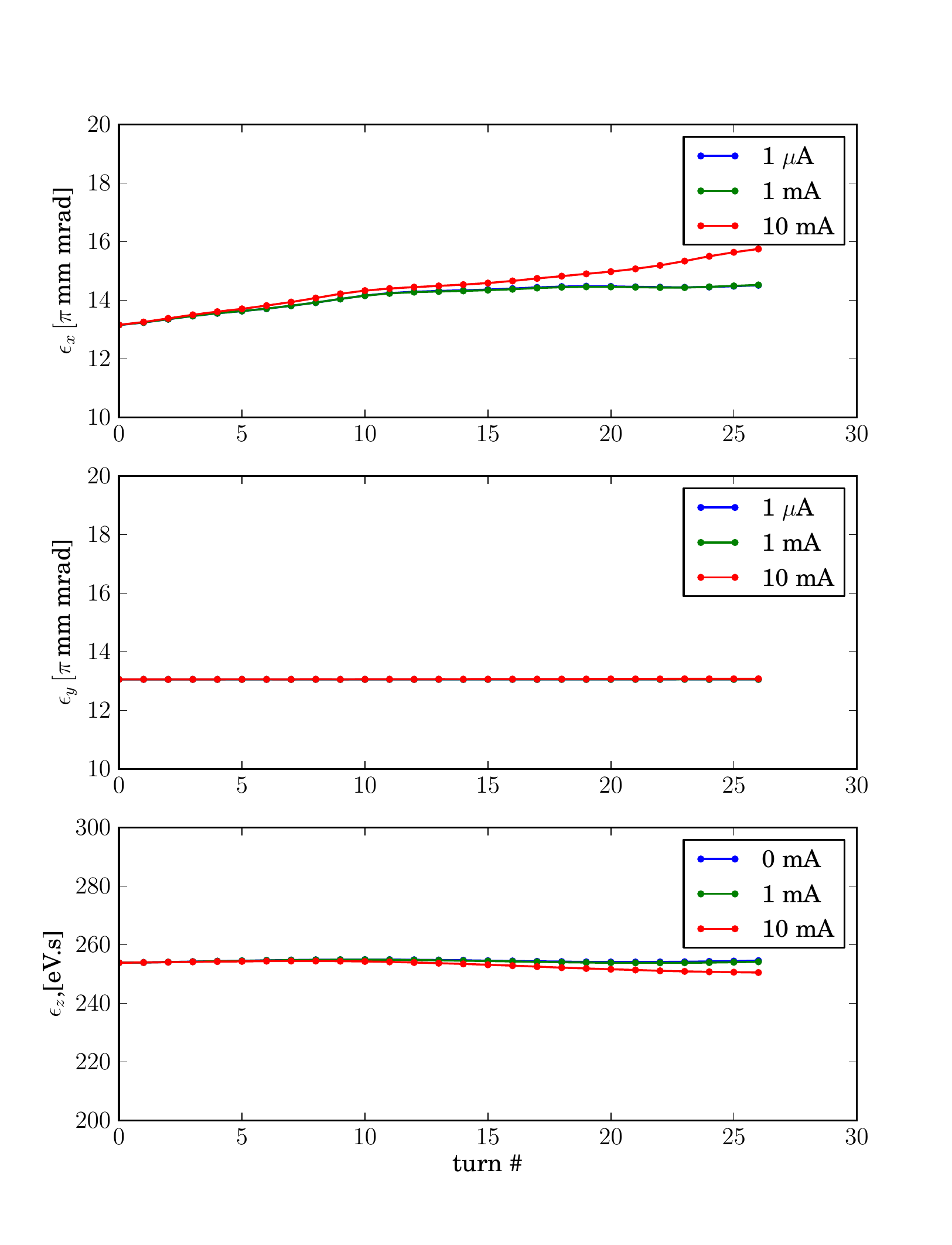}
\caption{Emittance variation at fixed energy at 610\,MeV for a 100\,mm long beam at varying average beam current.}
\end{figure}

\subsection{Simulations with acceleration}
\begin{figure}
\centering
\includegraphics[width=0.9\linewidth]{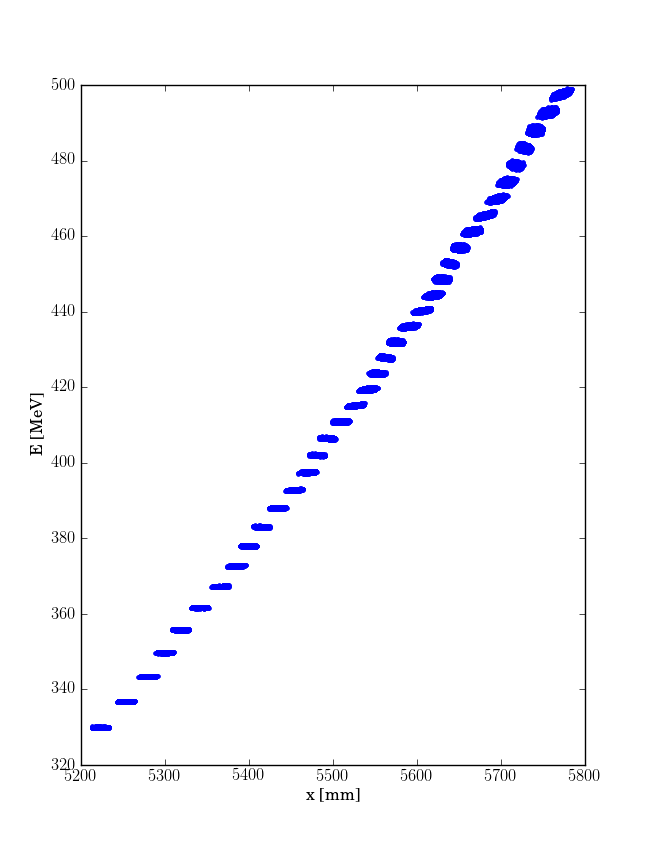}
\caption{Turn-by-turn radial position vs energy for a low intensity beam in the serpentine channel accelerated to 500\,MeV.\label{figure9}}
\end{figure}

Starting at 330\,MeV with 8\,MV/turn acceleration, a distribution is tracked until 500\,MeV. The simulation is stopped from reaching higher energies as the beam traverses a strong horizontal integer resonance $Q_{x}=2$ after 610\,MeV. In future the design should be optimised to avoid this resonance. The resulting distribution of radial position vs energy in Fig.~\ref{figure9} shows the turn-by-turn difference in radial position as the beam moves in the FFAG and the beginnings of the effects of distortion due to the serpentine channel are visible.

\section{Discussion}
In parallel with the lattice and beam dynamics studies outlined herein, work has been ongoing to design the magnets~\cite{weng13} and a possible superconducting RF system~\cite{johnstone-accapp} suitable for a nearly-isochronous ns-FFAG. 

Future work will concentrate on matching with space charge, adjusting and optimising the lattice parameters, studying injection and extraction mechanisms as well as conceptual engineering design of components.


\end{document}